\documentclass{ws-mpla}
\usepackage[T1]{fontenc}
\usepackage[latin9]{inputenc}
\usepackage{textcomp}
\usepackage{color}
\usepackage{amstext}
\usepackage{graphicx}
\usepackage{amssymb}

\newcommand{\noun}[1]{\textsc{#1}}

\begin{document}

\title{A topological approach to Neutrino masses by using exotic smoothness}

\author{Torsten Asselmeyer-Maluga}

\address{German Aerospace Center (DLR), Berlin, Germany and Copernicus Center
for Interdisciplinary Studies, ulica Szczepa\'{n}ska 1/5, 31-011 Krak\'ow,
Poland \\
 torsten.asselmeyer-maluga@dlr.de}

\author{Jerzy Kr{\'o}l}

\address{University of Silesia, Institute of Physics, Chorz\'ow, Poland and
Copernicus Center for Interdisciplinary Studies, ulica Szczepa\'{n}ska
1/5, 31-011 Krak\'ow, Poland \\
 iriking@wp.pl}

\maketitle
\pub{Received (Day Month Year)}{Revised (Day Month Year)} 
\begin{abstract}
In this paper, we will consider a cosmological model with two topological
transitions of the space. The smooth 4-dimensional spacetime of the
model admits topological transitions of its 3-dimensional slices.
The whole approach is inspired by a class of exotic smoothness structure
on $S^{3}\times\mathbb{R}$. In particular, this class of smoothness
structures induces two topological transitions. Then, we are able
to calculate the energy scales as associated to these topological
transitions. For the first transition we will get the value of the
GUT scale and the energy of the second transition is at the electroweak
scale. The topology of the exotic $S^{3}\times\mathbb{R}$ determines
both, the energy of the scales by certain topological invariants,
and the existence of the right-handed sterile neutrino. It is the
input for the seesaw mechanism. Secondly, based on this model, we
are able to calculate the neutrino masses which are in a very good
agreement with experiments. Finally, we will speculate, again based
on topology, why there are three generations of neutrinos and an asymmetry
between neutrinos and anti-neutrinos. 
\end{abstract}
\ccode{PACS Nos.: 14.60.Pq, 98.80.Cq, 14.60.St, 04.20.Gz, 02.40.Ma}

\section{Introduction}

The problem of fermion masses is one of the most fundamental issues
unresolved so far in the extensions of the standard model of elementary
particles {(SM, see the review \cite{Weinberg2018})}.
To understand the neutrino masses and the mixing of the neutrinos
is a particularly important ingredient of the problem. So far neutrinos
are the only fermions which are neutral with respect to the electric
as well color charge and they can admit masses. One of the accepted
mechanisms for explaining the small (Dirac) masses of neutrinos is
the seesaw mechanism (see for instance \cite{SeeSaw1977,SeeSaw1980,SeeSaw1980a}
and the recent paper \cite{SeeSaw2016} with a more complete bibliography)
which postulates a heavy right-handed Majorana neutrino.

Let us briefly discuss generalities of the simplest seesaw mechanism
of type I and how it fits in the proposed cosmological model based
on exotic $S^{3}\times\mathbb{R}$. In the minimal SM a neutrino is
represented by a Weyl spinor $\chi$ which belongs to the lepton isospin
doublet $L=\left(\begin{array}{c}
\chi\\
\psi
\end{array}\right)$. It is left-handed and massless, $\psi$ is the corresponding left-handed
charged lepton. In SM, there are 3 generations of leptons and each
of their neutrinos are represented by doublets as above.

Let now any left handed neutrino $\chi$ be a part of the doublet
$\left(\begin{array}{c}
\chi\\
\eta
\end{array}\right)$ where $\eta$ be a, postulated so far, right-handed neutrino (Weyl
spinor). It is a singlet under weak isospin and thus does not interact
weakly (sterile neutrino).

Possible mass terms are then generated by the following quadratic
form 
\[
(\chi,\eta)\left(\begin{array}{cc}
B' & M\\
M & B
\end{array}\right)\left(\begin{array}{c}
\chi\\
\eta
\end{array}\right).
\]
$B'$ has to be set to zero because it produces a nonrenormalizable
term. Thus the neutrino mass matrix reads 
\[
\left(\begin{array}{cc}
0 & M\\
M & B
\end{array}\right).
\]
However, $M$ is forced to vanish by symmetries of SM. The usual way
to introduce it will go through Yukawa-like interactions with the
Higgs field. These interactions will generate Dirac masses and $M$
can be non-zero. The value of $M$ is thus naturally of the order
of the vacuum expectation value of the Higgs field given by $246$
GeV or of the Higgs field mass of order $126$ GeV.

Now the value of $B$ is not fixed by the SM interactions since the
right-handed neutrinos are sterile and uncharged under any SM gauge
symmetry. Hence $B$ is a free parameter. Choosing $M\ll B$ the mass-matrix
eigenvalues read 
\[
\lambda_{1}\approx B\qquad\lambda_{2}\approx-\frac{M^{2}}{B}
\]
so that $\lambda_{1}$ is the mass of the right-handed neutrino and
$\lambda_{2}$ represents the mass of the left-handed neutrino. In
the unbroken phase one has $B=0$. But $B$ becomes large around the
scale of the spontaneous symmetry breaking via Yukawa interactions
with the Higgs field. An increase of the scale $B$ will induce smaller
masses of the left-handed neutrinos. Therefore one usually choose
the GUT scale of order $10^{15}$GeV.

The problem of this approach is the existence of two independent scales,
GUT and electroweak scale. The large size of $B$ is connected with
the GUT scale which guarantees that the Dirac masses of the left-handed
neutrinos are sufficiently small. But neutrinos carry no color and
electric charge, i.e. the next natural scale must be the electroweak
scale.

In this paper we will address the question whether the existence of
both scales can be uniquely determined by a cosmological model. We
show that indeed it can. The method of building the model is based
on the topological and geometric properties of the smooth cosmological
evolution represented by the exotic $S^{3}\times\mathbb{R}$ which
inherits the smoothness from the ambient exotic $R^{4}$. So thus
the scales appear as topologically supported quantities which fix
the value of the parameter $B$ into the GUT energy scale. Moreover,
the existence of the right-handed neutrino is also the effect of the
topology change. The realistic value of the Higgs mass is assigned
with the second topology change which indicates the electroweak scale.
Hence electroweak and GUT scales are not independent but rather determined
by topological properties of the model. {Mielke \cite{Mielke1977}
describes also a mechanism to generate the chirality of neutrinos
from topology by using the spacetime $S^{2}\times S^{1}\times\mathbb{R}$.
Interestingly, our model is connected to this work. The 3-manifolds
$Y_{k}$ in the infinite sequence $Y_{1}\to Y_{2}\to\ldots\to$ of
topology changes (see the next section) has the same homology as $S^{2}\times S^{1}$
(but not for the limit $Y_{\infty}$ which is a wildly embedded 3-sphere).
Therefore the whole process represents a spacetime which is homology
equivalent to $S^{2}\times S^{1}\times\mathbb{R}$ and the chirality
of neutrinos is generated by the same mechanism like in the model
of Mielke.}

The main part of the paper is the description of the topological constructions
showing how to translate (a part of) physics of SM into the intrinsic
topology of a 4-dimensional exotic spacetime. We show that fermions,
right-handed neutrinos, Higgs field and finally the GUT and electroweak
energy scales all this have their natural topological counterparts
in the cosmological model of our exotic $S^{3}\times\mathbb{R}$.
It is an amazing fact of this model that the numbers following from
the topology are realistic and serve as an input for the seesaw mechanism
to generate the neutrino masses.

\section{Basic model\label{sec:Basic-model}}

In this section we will describe the basic topological model of this
paper. The whole model was described completely in \cite{AsselmeyerKrol2018a}
where it was used to calculate the cosmological constant. In short,
the model postulates: 
\begin{itemize}
\item the spacetime is topologically $S^{3}\times\mathbb{R}$ but not smoothly 
\item there two transitions 
\[
S^{3}\stackrel{cork}{\longrightarrow}\Sigma(2,5,7)\stackrel{gluing}{\longrightarrow}P\#P\,.
\]
The first transition from the 3-sphere to the Brieskorn sphere 
\[
\Sigma(2,5,7)=\left\{ (x,y,z)\in\mathbb{C}^{3}\,|\, x^{2}+y^{5}+z^{7}=0,\,|x|^{2}+|y|^{2}+|z|^{2}=1\right\} 
\]
and a second transition from $\Sigma(2,5,7)$ to the sum of two Poincare
spheres $P\#P=(P\setminus D^{3})\cup_{S^{2}}(P\setminus D^{3})$ with
\[
P=\Sigma(2,3,5)=\left\{ (x,y,z)\in\mathbb{C}^{3}\,|\, x^{2}+y^{3}+z^{5}=0,\,|x|^{2}+|y|^{2}+|z|^{2}=1\right\}. 
\]

\end{itemize}
The model was completely described in \cite{AsselmeyerKrol2018a}
and partly in \cite{BizGom:96}. The relevant properties do not depend
on the particular smoothness structure of $S^{3}\times\mathbb{R}$
but the so-called standard smoothness structure is ruled out (there
is no topological transitions in this structure). The details of the
model will be described now but the reader can also skip the rest
of the section to keep only the two facts above in mind.

The distinguished feature of differential topology of manifolds in
dimension 4 is the existence of open 4-manifolds carrying a plenty
of non-diffeomorphic smooth structures. In the cosmological model
presented here, the special role is played by the topologically simplest
4-manifold, i.e. $\mathbb{R}^{4}$, which carries a continuum of infinitely
many different smoothness structures. Each of them except one, the
standard $\mathbb{R}^{4}$, is called \emph{exotic} $R^{4}$. All
exotic $R^{4}$ are Riemannian smooth open 4- manifolds homeomorphic
to $\mathbb{R}^{4}$ but non-diffeomorphic to the standard smooth
$\mathbb{R}^{4}$. The standard smoothness is distinguished by the
requirement that the topological product $\mathbb{R}\times\mathbb{R}^{3}$
is a smooth product. There exists only one (up to diffeomorphisms)
smoothing, the standard $\mathbb{R}^{4}$, where the product above
is smooth. There are two types of exotic $\mathbb{R}^{4}$: small
exotic $\mathbb{R}^{4}$ can be embedded into the standard $S^{4}$
whereas large exotic $\mathbb{R}^{4}$ cannot. In the following, an
exotic $\mathbb{R}^{4}$, presumably small if not stated differently,
will be denoted as $R^{4}$. In cosmology, one usually considers the
topology $S^{3}\times\mathbb{R}$ for the spacetime. But by using
the simple topological relations $\mathbb{R}^{4}\setminus D^{4}=S^{3}\times\mathbb{R}$
or $\mathbb{R}^{4}\setminus\left\{ 0\right\} =S^{3}\times\mathbb{R}$,
one obtains also an exotic $S^{3}\times\mathbb{R}$ from every exotic
$R^{4}$. In the following we will denote the exotic $S^{3}\times\mathbb{R}$
by $S^{3}\times_{\theta}\mathbb{R}$ to indicate the important fact
that there is no global splitting of $S^{3}\times_{\theta}\mathbb{R}$
into $S^{3}$-slices or it is not globally hyperbolic. This fact has
a tremendous impact on cosmology and therefore we will consider our
main hypothesis:\\
 \emph{MainHypo: The spacetime, seen as smooth four-dimensional manifold,
admits an exotic smoothness structure.}\\
 This hypothesis has the following consequences (see \cite{AsselmeyerKrol2018a}): 
\begin{itemize}
\item Any $R^{4}$ has necessarily non-vanishing Riemann curvature. Also
the $S^{3}\times_{\theta}\mathbb{R}$ has a non-vanishing curvature
along the $\mathbb{R}-$direction. 
\item Inside of $R^{4}$, there is a compact 4-dimensional submanifold $K\subset R^{4}$,
which is not surrounded by a smoothly embedded 3-sphere. Then there
is a chain of 3-submanifolds of $R^{4}$ $Y_{1}\to\cdots\to Y_{\infty}$
and the corresponding infinite chain of cobordisms 
\[
End(R^{4})=W(Y_{1},Y_{2})\cup_{Y_{2}}W(Y_{2},Y_{3})\cup\cdots
\]
where $W(Y_{k},Y_{k+1})$ denotes the cobordism between $Y_{k}$ and
$Y_{k+1}$ so that $R^{4}=K\cup_{Y_{1}}End(R^{4})$ where $\partial K=Y_{1}$.
The cobordism $W(Y_{k},Y_{k+1})$ is a 4-dimensional expression for
the topological transition $Y_{k}\to Y_{k+1}$. Furthermore one has
$End(R^{4})\subset S^{3}\times_{\theta}\mathbb{R}$. 
\item $R^{4}$ and $S^{3}\times_{\theta}\mathbb{R}$ embeds into the standard
$\mathbb{R}^{4}$ or $S^{4}$ but also in some other complicated 4-manifolds.
The construction of $R^{4}$ gives us a natural smooth embedding into
the compact 4-manifold $E(2)\#\overline{\mathbb{C}P^{2}}$ (with the
K3 surface $E(2)$) (see \cite{BizGom:96}). 
\end{itemize}
But every subset $K'$, $K'\subset K\subset R^{4}$, is surrounded
by a 3-sphere. This fact is the starting point of our model. Now we
choose a Planck-sized 3-sphere $S^{3}$ inside of the compact subset
$K\subset R^{4}$. This is the initial point where our cosmos starts
to evolve. By the construction of $R^{4}$, as mentioned above, there
exists the homology 3-sphere $\Sigma(2,5,7)$ inside of $K$ which
is the boundary of the Akbulut cork for $E(2)\#\overline{\mathbb{C}P^{2}}$.
(see chapter 9, \cite{GomSti:1999}). If $S^{3}$ is the starting
point of the cosmos as above, then $S^{3}\subset\Sigma(2,5,7)$. But
then we will obtain the first topological transition 
\[
S^{3}\to\Sigma(2,5,7)
\]
inside $R^{4}$. The construction of $R^{4}$ was based on the topological
structure of $E(2)$ (the K3 surface). $E(2)$ splits topologically
into a 4-manifold $|E_{8}\oplus E_{8}|$ with intersection form $E_{8}\oplus E_{8}$
(see \cite{GomSti:1999}) and the sum of three copies of $S^{2}\times S^{2}$.
{Here, $E_{8}$ denotes a $8\times8$ matrix of determinant
$1$ with integer entries. It is the Cartan matrix of the Lie algebra
$e_{8}$ of the exceptional Lie group $E_{8}$. Now $E_{8}$ denotes
the matrix of the intersection form (as integer
bilinear form) of certain 4-manifold $|E_{8}|$ (with a boundary - see below). }

In the topological splitting of $E(2)$ 
\begin{equation}
|E_{8}\oplus E_{8}|\times\underbrace{\left(S^{2}\times S^{2}\right)\times\left(S^{2}\times S^{2}\right)\times\left(S^{2}\times S^{2}\right)}_{3\left(S^{2}\times S^{2}\right)}\label{eq:splitting-K3}
\end{equation}
the 4-manifold $|E_{8}\oplus E_{8}|$ has a boundary which is the
sum of two Poincare spheres $P\#P$. Here we used the fact that a
smooth 4-manifold with intersection form $E_{8}$ must have a boundary
(which is the Poincare sphere $P$), otherwise it would contradict
Donaldson's theorem. Then any closed version of $|E_{8}\oplus E_{8}|$
does not exist and this fact is the reason for the existence of exotic
$R^{4}$. To express it differently, the $R^{4}$ lies between this
3-manifold $\Sigma(2,5,7)$ and the sum of two Poincare spheres $P\#P$.
We analyzed this spacetime in \cite{AsselmeyerKrol2012}. It is interesting
to note that \emph{the number of the $S^{2}\times S^{2}$ components
must be three or more otherwise the corresponding spacetime is not
smooth!}

Therefore we have two topological transitions resulting from the embedding
into $E(2)\#\overline{\mathbb{C}P^{2}}$ 
\[
S^{3}\stackrel{cork}{\longrightarrow}\Sigma(2,5,7)\stackrel{gluing}{\longrightarrow}P\#P\,.
\]
These two topological transitions constitute the main idea of this work together with the mapping of them into two
different energy scales: 
\begin{itemize}
\item $S^{3}\to\Sigma(2,5,7)$ which must be related to the GUT scale in between
$10^{4}$ to $10^{3}$ times lower then the Planck scale, and 
\item $\Sigma(2,5,7)\to P\#P$ related to the electroweak scale $246$ GeV
or to the Higgs mass $126$ GeV. 
\end{itemize}
We will show in the next section that the energy scales of the transitions
are consistent with this idea.

\section{Energy scales and topological transitions\label{sec:Energy-scales}}

In this section we will show how the change of the energy scale is
driven by the topological transitions 
\[
S^{3}\stackrel{cork}{\longrightarrow}\Sigma(2,5,7)\stackrel{gluing}{\longrightarrow}P\#P\,.
\]
Both transitions have different topological descriptions. The first
transition $S^{3}\to\Sigma(2,5,7)$ can be realized by a smooth cobordism,
i.e. by a smooth 4-manifold $M$ with boundary $\partial M=S^{3}\sqcup\Sigma(2,5,7)$.
But dimension 4 is special and one needs an infinite process to realize
this transition, called a Casson handle. In \cite{AsselmeyerKrol2018a}
we described this situation extensively. In particular, we obtained
a scaling formula between the length scale $a_{0}$ of the 3-sphere
$S^{3}$ and the length scale $a$ of $\Sigma(2,5,7)$: 
\begin{equation}
a=a_{0}\cdot\exp\left(\frac{3}{2\cdot CS(\Sigma(2,5,7))}\right)\label{eq:length-scaling}
\end{equation}
where $CS(\Sigma(2,5,7))$ is the Chern-Simons invariant of this manifold.
{The Chern-Simons invariant $CS(N)$ of a 3-manifold
$N$ depends on the representation of the fundamental group $\pi_{1}(N)$
into $SU(2)$ or $SL(2,\mathbb{C})$, see the paper \cite{KirKla:90}.
The minimal value for all representations is an invariant which will
be exclusively considered in the following. By using the solution
to the 3D Poincare conjecture, the value $CS(N)=0$ is only possible
for the 3-sphere $N=S^{3}$. For this value, the expression above
for $a$ would diverge but the 3-sphere cannot appear in the formula
for $a$ otherwise it contradicts the appearance of an exotic smoothness
structure and a hyperbolic structure used in the derivation of this
formula. In quantum gravity, the Chern-Simons invariant was used in
a special class of states, the Kodama state \cite{Kodama1990} used
in Loop quantum gravity. It is known to be unphysical for a variety
of reasons as discussed by Witten \cite{Witten2003}, but it is still
interesting to understand what Kodama state and CS invariants could correspond to in quantum gravity. The Chern-Simons term was considered
also by Mielke \cite{Mielke1998} in the context of the teleparallelism theory of gravity, $GR_{||}$, which is a gauge theory of translations. Then the Chern-Simons invariant
is the solution of the teleparallel Ashtekar constraints \cite{Mielke2002}. In all cases, the Chern-Simons term
is not a denominator of a fraction. In the presented approach, hyperbolic geometry explains how the CS term appears in the above formula (\ref{eq:length-scaling}).}

In \cite{AsselmeyerKrol2014}, we also derived this formula by relating
it to the levels of the tree representing the Casson handle. By using
the shortening 
\[
\vartheta=\frac{3}{2\cdot CS(\Sigma(2,5,7))}
\]
we obtained 
\[
a=a_{0}\cdot\sum_{n=0}^{\infty}\frac{\vartheta^{n}}{n!}
\]
or the $n$th level will contribute by the term $\frac{\vartheta^{n}}{n!}$
to the change of the length scale. The Casson handle is directly related
to the topology change. Therefore, it is natural from the physics
point of view to identify the level number $n$ with the time coordinate.
Now we will determine the smallest possible time change $\triangle t$. {This
change can be understood as introducing an uncertainty $\triangle t$
which results in an uncertainty of the energy $\triangle E$, see
the discussion of Hilgevoord\cite{time-energy-uncertain1998} supporting
our view.} Then we are able to determine the corresponding energy
change $\triangle E$. This is given by the well-known
relation $\Delta t\sim h/\Delta E$ which follows from the quantum nature of fluctuations where the smallest time change gives rise to the corresponding energy
change.

Therefore, due to the quantum nature of the fluctuations we are enforced to determine the shortest time change. But this change has natural topological interpretation via
the minimal number of levels in the tree of the Casson handle where the topology change appears. How many levels are necessary to get such topology change?
In \cite{Fre:88} Freedman answered this question: three levels are
needed to embed entire CH in the 3-tower of a Casson handle! Then if we assume that the shortest time scale of one level
is given by the Planck time $t_{Planck}$ then we will get for the
smallest time change 
\begin{equation}
\Delta t=\left(1+\vartheta+\frac{\vartheta^{2}}{2}+\frac{\vartheta^{3}}{6}\right)t_{Planck}\label{eq:time-change-firt}
\end{equation}
so that we obtain for the energy scale 
\begin{equation}
\Delta E=\frac{E_{Planck}}{1+\vartheta+\frac{\vartheta^{2}}{2}+\frac{\vartheta^{3}}{6}}\quad.\label{eq:GUT-scale}
\end{equation}
For the transition $S^{3}\to\Sigma(2,5,7)$ we have the values 
\[
CS(\Sigma(2,5,7))=\frac{9}{280},\quad\vartheta=\frac{140}{3},\quad\Delta E_{GUT}\approx0.67\cdot10^{15}\, GeV
\]
and thus the first energy scale is of an order 
\[
B\approx10^{15}GeV
\]
of the GUT energy scale.

{In the formula (\ref{eq:time-change-firt}) above,
we obatined a relation between time and the Chern-Simons invariant.
This relation has some similarities with a proposal of Smolin and
coworkers\cite{SmolinSoo1995} which has been discussed recently\cite{Smolin-CS-time2018}.
In contrast to our formula, the time is directly given by the Chern-Simons
invariant (or its imaginary part). The difference to our proposal
can be explained by the essential role played by hyperbolic geometry in the derivation
of (\ref{eq:time-change-firt}) along with a (dynamically induced) time foliation
which is fixed by rigidity in hyperbolic geometry.}

As described in \cite{AsselmeyerKrol2018a}, for the second transition
$\Sigma(2,5,7)\to P\#P$, we need a different argumentation. As shown
in \cite{AsselmeyerKrol2012}, this transition cannot be represented
by a smooth cobordism (a smooth 4-manifold with boundary $P\#P\sqcup\Sigma(2,5,7)$).
But we are able to get a smooth 4-manifold by adding three hyperbolic
3-manifolds (see \cite{AsselmeyerKrol2012} for an explanation). Here
we will need entire tower with all infinite many levels of Casson handles to realize the transition, which stays
in contrast to the 3-stages of the previous transition. As argued in \cite{AsselmeyerBrans2015} these hyperbolic 3-manifolds can
be physically interpreted as matter.

But adding of three hyperbolic 3-manifolds means that we need now three different Casson handles for the transition. To express
it differently: From the physics point of view, we have three channels
for the change $\Sigma(2,5,7)\to P\#P$. The time change $\triangle t$
of this change will be determined by one channel, in contrast to the
scaling of the length where one needs the whole expression (see \cite{AsselmeyerKrol2018a}).
Therefore we have to modify the usual exponent by the number of channels
\[
\frac{1}{3}\left(\frac{3}{2\cdot CS(P\#P)}\right)
\]
to obtain the change in the time scale 
\[
\Delta t=\tilde{\triangle t}\cdot\exp\left(\frac{1}{2\cdot CS(P\#P)}\right)
\]
where $\tilde{\triangle t}$ represents the time change for one level
in the Casson handle, see above. Then with the value 
\[
CS(P\#P)=\frac{1}{60}
\]
we will get for the time change of both transitions 
\[
\triangle t=\frac{t_{Planck}\cdot\exp\left(-\frac{1}{2\cdot CS(P\#P)}\right)}{1+\vartheta+\frac{\vartheta^{2}}{2}+\frac{\vartheta^{3}}{6}}
\]
where we used (\ref{eq:time-change-firt}). Then we obtain for
the corresponding energy scale 
\begin{equation}
M=\Delta E=\frac{E_{Planck}\cdot\exp\left(-\frac{1}{2\cdot CS(P\#P)}\right)}{1+\vartheta+\frac{\vartheta^{2}}{2}+\frac{\vartheta^{3}}{6}}\approx63GeV\quad\label{eq:Higgs-scale}
\end{equation}
which is exactly half of the Higgs mass and we will identify this
value with the second energy scale. Now we have fixed the two energy
scales for the seesaw mechanism. But one ingredient is missing: which
structures describes the usual left-handed neutrino in our topological scheme and how did we get the very massive right-handed neutrino (sterile neutrino). This
question will be addressed in the next section.

\section{Fermions and knot complements\label{sec:Fermions-and-knot}}

In this section we will discuss the topological reasons for the existence
of fermions. In our previous paper \cite{AsselmeyerBrans2015} we
obtained a relation between an embedded 3-manifold and a spinor in
the spacetime. The main idea can be simply described by the following
line of argumentation. Let $\iota:\Sigma\hookrightarrow M$ be an
embedding of the 3-manifold $\Sigma$ into the 4-manifold $M$ with
the normal vector $\vec{N}$. A small neighborhood $U_{\epsilon}$
of $\iota(\Sigma)\subset M$ looks like $U_{\epsilon}=\iota(\Sigma)\times[0,\epsilon]$.
Furthermore we identify $\Sigma$ and $\iota(\Sigma)$ ($\iota$ is
an embedding). Every 3-manifold admits a spin structure with a \noun{spin
bundle}, i.e. a principal $Spin(3)=SU(2)$ bundle (spin bundle) as
a lift of the frame bundle (principal $SO(3)$ bundle associated to
the tangent bundle). There is a (complex) vector bundle associated
to the spin bundle (by a representation of the spin group), called
\noun{spinor bundle} $S_{\Sigma}$, {see Trautman \cite{Trautman2005}
for a careful definition of spinors}. A section in the spinor bundle
is called a spinor field (or a spinor). In case of a 4-manifold, we
have to assume the existence of a spin structure. But for a manifold
like $M=S^{3}\times\mathbb{R}$, there is no restriction, i.e. there
is always a spin structure and a spinor bundle $S_{M}$. In general,
the unitary representation of the spin group in $D$ dimensions is
$2^{[D/2]}$-dimensional. From the representational point of view,
a spinor in 4 dimensions is a pair of spinors in dimension 3. Therefore,
the spinor bundle $S_{M}$ of the 4-manifold splits into two sub-bundles
$S_{M}^{\pm}$ where one sub-bundle, say $S_{M}^{+},$ can be related
to the spinor bundle $S_{\Sigma}$ of the 3-manifold. Then the spinor
bundles are related by $S_{\Sigma}=\iota^{*}S_{M}^{+}$ with the same
relation $\phi=\iota_{*}\Phi$ for the spinors ($\phi\in\Gamma(S_{\Sigma})$
and $\Phi\in\Gamma(S_{M}^{+})$). Let $\nabla_{X}^{M},\nabla_{X}^{\Sigma}$
be the covariant derivatives in the spinor bundles along a vector
field $X$ as section of the bundle $T\Sigma$. Then we have the formula
\begin{equation}
\nabla_{X}^{M}(\Phi)=\nabla_{X}^{\Sigma}\phi-\frac{1}{2}(\nabla_{X}\vec{N})\cdot\vec{N}\cdot\phi\label{eq:covariant-derivative-immersion}
\end{equation}
with the embedding $\phi\mapsto\left(\begin{array}{c}
0\\
\phi
\end{array}\right)=\Phi$ of the spinor spaces from the relation $\phi=\iota_{*}\Phi$. There are, certainly, two possible embeddings. For later
use we will choose the left-handed version. The expression $\nabla_{X}\vec{N}$
is the second fundamental form of the embedding where the trace $tr(\nabla_{X}\vec{N})=2H$
is related to the mean curvature $H$. Then from (\ref{eq:covariant-derivative-immersion})
one obtains the following relation between the corresponding Dirac
operators 
\begin{equation}
D^{M}\Phi=D^{\Sigma}\phi-H\phi\label{eq:relation-Dirac-3D-4D}
\end{equation}
with the Dirac operator $D^{\Sigma}$ on the 3-manifold $\Sigma$.
This relation (as well as (\ref{eq:covariant-derivative-immersion}))
is only true for the small neighborhood $U_{\epsilon}$ where the
normal vector points is parallel to the vector defined by the coordinates
of the interval $[0,\epsilon]$ in $U_{\epsilon}$. In \cite{AsselmeyerRose2012},
we extend the spinor representation of an immersed surface into the
3-space to the immersion of a 3-manifold into a 4-manifold according
to the work in \cite{Friedrich1998}. Then the spinor $\phi$ defines
directly the embedding (via an integral representation) of the 3-manifold.
Then the restricted spinor $\Phi|_{\Sigma}=\phi$ is parallel transported
along the normal vector and $\Phi$ is constant along the normal direction
(reflecting the product structure of $U_{\epsilon}$). But then the
spinor $\Phi$ has to fulfill 
\begin{equation}
D^{M}\Phi=0\label{eq:Dirac-equation-4D}
\end{equation}
in $U_{\epsilon}$ i.e. $\Phi$ is a parallel spinor. Finally we get
\begin{equation}
D^{\Sigma}\phi=H\phi\label{eq:Dirac3D-mean-curvature}
\end{equation}
with the extra condition $|\phi|^{2}=const.$ (see \cite{Friedrich1998}
for the explicit construction of the spinor with $|\phi|^{2}=const.$
from the restriction of $\Phi$). The idea of the paper \cite{AsselmeyerBrans2015}
was to use the Einstein-Hilbert action for a spacetime with
boundary $\Sigma$. The boundary term is the integral of the mean
curvature for the boundary, see \cite{Ashtekar08,Ashtekar08a}. Then
by the relation (\ref{eq:Dirac3D-mean-curvature}) we will obtain
\begin{equation}
\intop_{\Sigma}H\,\sqrt{h}\, d^{3}x=\intop_{\Sigma}\bar{\phi}\, D^{\Sigma}\phi\,\sqrt{h}d^{3}x\label{eq:relation-mean-curvature-action-to-dirac-action}
\end{equation}
using $|\phi|^{2}=const.$ As shown in \cite{AsselmeyerBrans2015},
the extension of the spinor $\phi$ to the 4-dimensional spinor $\Phi$
by using the embedding 
\begin{equation}
\Phi=\left(\begin{array}{c}
0\\
\phi
\end{array}\right)\label{eq:embedding-spinor-3D-4D}
\end{equation}
can be seen as embedding, if (and only if) the 4-dimensional
Dirac equation 
\begin{equation}
D^{M}\Phi=0\label{eq:4D-Dirac-equation}
\end{equation}
on $M$ is fulfilled (using relation (\ref{eq:relation-Dirac-3D-4D})).
This Dirac equation is obtained by varying the action 
\begin{equation}
\delta\intop_{M}\bar{\Phi}D^{M}\Phi\sqrt{g}\: d^{4}x=0\label{eq:4D-variation}
\end{equation}
In \cite{AsselmeyerBrans2015} we went a step further and discussed
the topology of the 3-manifold leading to a fermion. On general grounds,
one can show that a fermion is given by a knot complement admitting
a hyperbolic structure. The connection between the knot and the particle
properties is currently under investigation. But first calculations
seem to imply that the particular knot is only important for the dynamical
state (like the energy or momentum) but not for charges, flavors etc.
But some properties can be derived from the approach above: 
\begin{itemize}
\item The construction of the 4-spinor $\Phi$ by (\ref{eq:embedding-spinor-3D-4D})
gives us a fermion of fixed chirality. Currently we know only one
such particle, the neutrino. 
\item A Dirac operator over $\Sigma$ is strongly related to this manifold,
i.e. a local diffeomorphism cannot change the operator. But there
are diffeomorphisms which are not connected to the identity or the
diffeomorphism group is not connected. The mapping class group, the
group of connection components for the diffeomorphism group, labels
the different Dirac operators on $\Sigma$ (see also \cite{AsselmeyerRose2012}). 
\end{itemize}
Even the last point has a strong impact on the model. Now we will
consider the case that the diffeomorphism group $Diff(\Sigma)$ has
two disjoint components, i.e. $\pi_{0}(Diff(\Sigma))=\mathbb{Z}_{2}$.
The group $\pi_{0}(Diff(\Sigma))$ is known as mapping class group.
These two different components lead to two different Dirac operators on the 3-manifold
$\Sigma$ which can be combined in a single 4-dimensional Dirac operator.
In case of the neutrino, one would obtain a left-handed and a right-handed
neutrino. Keeping this fact in mind, we will look at the first transition
\[
S^{3}\to\Sigma(2,5,7)
\]
and the mapping class is known to be

\begin{equation}
\pi_{0}(Diff(\Sigma(2,5,7)))=\mathbb{Z}_{2}\:.\label{eq:MCG-Brieskorn}
\end{equation}
Interestingly, the second transition 
\[
\Sigma(2,5,7)\to P\#P
\]
leads to $P\#P$ having a trivial mapping class group 
\begin{equation}
\pi_{0}(Diff(P))=1\:.\label{eq:MCG-Poincare}
\end{equation}
{Here, we recommend the papers \cite{HatcherWahl2010-mcg,McCulloughHong2013-mcg}
for the relevant informations about mapping class groups of 3-manifolds.}
Therefore we have the following picture: 
\begin{itemize}
\item At the end of the first transition $S^{3}\to\Sigma(2,5,7)$, there
is a left-handed neutrino and a right-handed neutrino. 
\item The right-handed neutrino is caused by the first transition. Therefore
the mass of the right-handed neutrino must be proportional to the
energy scale of the first transition which is the GUT scale (see \ref{eq:GUT-scale}). 
\item The left-handed neutrino is not connected with any special transition.
Then the corresponding mass must be (much) smaller. 
\end{itemize}
For the calculation of this small mass we will use the seesaw mechanism
as described in the next section.

\section{Seesaw mechanism \label{sec:Seesaw-mechanism-Higgs}}

Now it is time to join the topological approach with the seesaw mechanism
to generate the neutrino mass. So, let us present an overview of this
approach 
\begin{itemize}
\item Fermions are given by hyperbolic knot complements represented by a
3D spinor. The corresponding 4D spinor is chiral and given by the
embedding (\ref{eq:embedding-spinor-3D-4D}). The corresponding particle,
the neutrino, must be left-handed. 
\item In the first transition, $S^{3}\to\Sigma(2,5,7)$, a right-handed
neutrino is generated because of a nontrivial mapping class group
(\ref{eq:MCG-Brieskorn}). It is strongly connected with this transition.
The energy scale is given by (\ref{eq:GUT-scale}) which gives $B\approx0.67\cdot10^{15}GeV$. 
\item In the second transition, $\Sigma(2,5,7)\to P\#P$, we have only the
left-handed neutrino. The energy scale can be expressed by (\ref{eq:Higgs-scale})
giving approximately $M\approx63GeV$. 
\end{itemize}
Now we have all ingredients to calculate the neutrino mass by the
seesaw mechanism. We can start with the non-diagonal mass matrix like
in the Introduction 
\[
\left(\begin{array}{cc}
0 & M\\
M & B
\end{array}\right)
\]
with two mass scales $B$ and $M$ fulfilling $M\ll B$. This matrix
has eigenvalues 
\[
\lambda_{1}\approx B\qquad\lambda_{2}\approx-\frac{M^{2}}{B}
\]
so that $\lambda_{1}$ is the mass of the right-handed neutrino and
$\lambda_{2}$ represents the mass of the left-handed neutrino. Above
we fix the scales to the values (\ref{eq:GUT-scale}) and (\ref{eq:Higgs-scale})
\[
B\approx0.67\cdot10^{15}GeV,\quad M\approx63GeV
\]
and we will obtain for the neutrino mass 
\[
m=\frac{M^{2}}{B}\approx0.006eV
\]
{The sum of the three neutrino masses is constrained
by the results of the PLANCK mission \cite{PlanckCosmoParameters2013,PlanckCosmoParameters2015}.
If we assume three identical masses we will get for the sum $0.018$eV.
The PLANCK mission determined only an upper value of $0.3$eV but
the Baryon Acoustic Oscillations (BAO) lower this value to $0.12-0.17$eV
for the sum in good agreement with our result. There are many experimental
results\cite{Neutrino-mass2015,Neutrino-mass2016} for the neutrino
masses obtained with different methods (CMB, BAO, weak lensing). The
smallest upper bound of $0.06$eV can be found in the work \cite{Neutrino-mass-KmLAND-Zen2016}.
All experimental results give only upper bounds and one has to wait for farther results showing how small the neutrino masses could be. The model presented here demonstrates how to generate small masses in a natural way.}

\section{Conclusion}

We were able to get the seesaw mechanism for generating neutrino masses
in a topological model for the cosmic evolution based on two topological
transitions. This model describes the cosmology of the evolving universe
and is based on the exotic smoothness of $S^{3}\times_{\Theta}\mathbb{R}\subset R^{4}$.
There are two topology changes within the 3-dimensional slice of $S^{3}\times_{\Theta}\mathbb{R}$,
i.e. $S^{3}\to\Sigma(2,5,7)$ related to the GUT scale, and $\Sigma(2,5,7)\to P\#P$
related to the electroweak scale. Both scales can be calculated by
using the topological numbers of the transitions rather than just
assigned it from the outside. The basic ingredient of the seesaw mechanism,
i.e. the right-handed neutrino, is also topologically supported in
the model. It is connected with the nontrivial mapping class group
$\pi_{0}(Diff(\Sigma(2,5,7)))=\mathbb{Z}_{2}$ of the first transition.
On the other hand the mapping class $\pi_{0}(Diff(P))=1$ of the second
transition is trivial. These topological results determine much of
the physics behind the seesaw mechanism which was demonstrated in
our model. {Again, we have to state that the results
depend strongly on the usual incorporation of the Chern-Simons invariant.
In this paper we used the expression (\ref{eq:length-scaling}) with
the Chern-Simons invariant at the denominator. Usually, the Chern-Simons
invariant appeared in the numerator like in the Kodama state \cite{Kodama1990},
in teleparallel gravity as the solution of the constraints \cite{Mielke2002}
or to define time \cite{SmolinSoo1995,Smolin-CS-time2018}. But the
expression (\ref{eq:length-scaling}) is crucial to get the correct
results for the energy scales and neutrino masses.}

{We can speculate about farther consequences of the model.}
As explained above, the splitting (\ref{eq:splitting-K3}) has a tremendous
impact on the spacetime, the exotic $S^{3}\times_{\theta}\mathbb{R}$.
The appearance of three $S^{2}\times S^{2}$ has an influence on the
energy scale (see section \ref{sec:Energy-scales}). In section \ref{sec:Fermions-and-knot},
we constructed the fermion as embedded 3-manifolds (knot complements).
As shown in \cite{AsselmeyerBrans2015}, the particular Casson handle
determines the fermion. Therefore the three $S^{2}\times S^{2}$ in
(\ref{eq:splitting-K3}) give three Casson handles with three kinds
of fermions. To express it differently, \emph{there are three generations}.
Currently we have no idea about the difference between the generations
but as discussed above and in \cite{AsselmeyerKrol2012}, there must
exist at least three generations. {Otherwise the spacetime
cannot be smooth.} It is interesting to note that the three $S^{2}\times S^{2}$
cannot be distinguished at the level of spacetime. Therefore, the
whole approach must be invariant with respect to the symmetric group
$S_{3}$ isomorphic to the dihedral group $D_{3}=\langle r,s\,|\, r^{3}=e=s^{2},srs=r^{-1}\rangle$.
We conjecture that this symmetry should appear somehow in the mixing
matrix of the neutrinos (Pontecorvo-Maki-Nakagawa-Sakata matrix).
Finally, the topology of the spacetime opens also a way to produce
an asymmetry between neutrinos and anti-neutrinos. In section \ref{sec:Fermions-and-knot},
we described the appearance of the fermion. We constructed a Dirac
operator $D$ on the spacetime. The kernel of this operator $\ker D$
are the solutions of the Dirac equation which correspond to the neutrino.
The kernel $\ker D^{\dagger}$ of the conjugated operator $D^{\dagger}$
is associated to the anti-neutrino. Thanks to the Atiyah-Singer index
theorem (see \cite{Nak:89,Nas:91}) there is a relation between the
difference $\dim\ker D-\dim\ker D^{\dagger}$ and the topology of
the underlying manifold or this difference is given by a topological
invariant (so-called A-roof genus). In our case, one obtains 
\[
\dim\ker D-\dim\ker D^{\dagger}=2
\]
and one has an asymmetry between the neutrinos and anti-neutrinos,
because the number of possible solutions are different. Interestingly,
the difference is given by the number of $E_{8}$ factors in the splitting
(\ref{eq:splitting-K3}) of $E(2)\#\overline{\mathbb{C}P^{2}}$. Currently
we have no idea to quantify this difference to get in contact with
current measurements.

As we have already mentioned many details must be worked out. It is
ongoing work on this topic. The entire approach relating topology
in dimension 3 and 4 and physics of SM is by no means exhausted, it
is rather at its initial stage. It is surprising indeed how direct
and specific connections of topology and physics can be worked out
and how many connections are presumably hidden.

\section*{Acknowledgement}

This publication was made possible through the support of a grant
from the John Templeton Foundation (Grant No. 60671),



\end{document}